\documentclass[twocolumn, pre]{revtex4}

\usepackage{amssymb}
\usepackage{amsmath}
\usepackage{graphics}
\usepackage{graphicx}

\newcommand{\be}{\begin{equation}}
\newcommand{\ee}{\end{equation}}

\begin{document}

\title{Ionic conductivity on a wetting surface}
\date{\today}
\author{Brian Skinner}
\author{M.S. Loth}
\author{B.I. Shklovksii}
\affiliation{Theoretical Physics Institute, University of Minnesota, Minneapolis, Minnesota 55455}

\begin{abstract}
Recent experiments measuring the electrical conductivity of DNA molecules highlight the need for a theoretical model of ion transport along a charged surface.  Here we present a simple theory based on the idea of unbinding of ion pairs.  The strong humidity dependence of conductivity is explained by the decrease in the electrostatic self-energy of a separated pair when a layer of water (with high dielectric constant) is adsorbed to the surface.  We compare our prediction for conductivity to experiment, and discuss the limits of its applicability.
\end{abstract} \maketitle

\section{Introduction}

For materials where electrical conduction occurs at the surface, the adsorption of water molecules can have a very strong effect on the conductivity.  A variety of experiments over the last four decades have examined the effect of water adsorption on surface conductivity for materials like quartz \cite{Awakuni}, silica gel \cite{Anderson}, polymer films \cite{Sakai}, lipid membranes \cite{Heim}, and various ceramics \cite{Sadaoka, Seiyama}. In these materials the conductivity changes by as much as six orders of magnitude as the atmospheric humidity is varied, increasing with humidity in most cases by a simple exponential.  This dependence has been exploited to design humidity detectors from ceramic materials \cite{Seiyama}.

In recent years, studies examining the electrical properties of double-helical DNA and DNA bundles have found a similar dependence of conductivity on relative humidity \cite{HanHa, Yamahata, Tuukkanen}.  Experimental data shows that the conductivity of DNA increases by a simple exponential over nearly the entire range of relative humidity ($10\%$ to $100\%$), and spans roughly six orders of magnitude.  Experiments on flat DNA films show a similar exponential dependence \cite{Kleine-Ostmann, Otsuka, Matsuo, Armitage}.  While the mechanism for electrical conduction in DNA remains controversial, it is widely believed to occur through the movement of ions along the outer surface of the double-helix \cite{Kleine-Ostmann, Laudat, Brovchenko}.

The similarity between conductivity measurements for DNA and for other surface-conducting materials suggests that there may be some universal law explaining the exponential dependence of conductivity on relative humidity.  As of yet, there is no generally accepted theory of electrical conduction on a wetting surface.  In this paper we propose such a theory, based on the idea of unbinding of ion pairs.

\section{Adsorption of the water layer}

In this paper we examine the case of a uniform, completely-wetting surface (contact angle $\theta = 0$).  On such a surface we can consider the adsorbed water to form a complete, flat layer rather than a set of isolated droplets.  The case of small but finite contact angle $0 < \theta \ll 1$ (incomplete wetting) is discussed briefly at the end of this section.

There are a number of theoretical treatments which yield estimates for the water layer thickness as a function of vapor pressure $p$.  The most cited of these is the so-called BET formula \cite{BET}, which has been widely employed to describe the adsorption of the first monolayer of a gas onto a substrate.  Despite its practicality, the theoretical value of the BET formula for thicker films remains questionable.  Its description of the water surface as a ``molecular pile" ignores the effects of surface tension, and it is known in many cases to predict significantly greater adsorption at high humidities than is actually observed \cite{Hill, Teller}.

The most general way of calculating the water layer thickness is by defining the chemical potential $\mu = \mu' + \mu_l$ of a water molecule within the layer, where $\mu_l$ is the chemical potential of bulk water and $\mu'$ is the component of chemical potential resulting from interactions with the solid adsorbent.  In general $\mu'$ varies with the distance $z$ from the solid surface such that $\mu'(z \rightarrow \infty) = 0$.  The equilibrium water layer thickness is the distance $d$ at which $\mu'(z = d)$ is equal to the free energy of condensation for a single water molecule \cite{Landau-stat}:
\be 
F_{ads} = k_B T \ln(p/p_0) \label{eq:Fads} .
\ee
Here, $p_0$ is the saturated vapor pressure and $k_BT$ is the thermal energy.

The simplest way to estimate $\mu'$ is to assume that it is the result of Van der Waals forces between the water molecule and the adsorbent.  If we assume the adsorbent to occupy the half-space $z < 0$, then we can estimate the chemical potential $\mu'(z)$ of a water molecule at $z > 0$ to be \cite{colloidal}
\be 
\mu'(z) = w^3 \frac{H}{6 \pi z^3} . \label{eq:mu'} 
\ee
Here, $w^3$ is taken to mean the volume occupied by a single water molecule in the liquid phase ($w \approx 3.1$ \r{A}) and $H$ is the (negative) Hamaker constant for the water-adsorbent interaction \cite{Hfootnote}.  For water-DNA interaction, the Hamaker constant can be estimated as $H \approx -0.4$ eV \cite{Israelachvili, Zhang}.  The Van der Waals energy of Eq$.$ (\ref{eq:mu'}) is considered a source of ``disjoining pressure", meaning that it works to increase the thickness of the water layer in opposition to the condensation free energy of Eq$.$ (\ref{eq:Fads}).  The resulting water layer thickness satisfies $\mu'(z = d) = F_{ads}$, or
\be 
d = w \left( \frac{H}{6 \pi k_B T \ln(p/p_0)} \right)^{1/3} \label{eq:d-RH},
\ee
a result first derived by Frenkel \cite{Frenkel}.  For some materials this prediction is in good agreement with adsorption experiments \cite{Hill, Teller}, while for other materials there are important ``structural" contributions to the disjoining pressure associated with crystalline-like ordering of water molecules \cite{Derjaguin2, Derjaguin}.  These ``structural" forces are highly dependent on the molecular nature of the substrate surface, and experiments probing their significance for materials like quartz have produced contradictory results \cite{Pashleyquartz}.  For the remainder of this paper we will assume that structural forces are unimportant.

In the case of an incompletely wetting surface with contact angle $0 < \theta \ll 1$, the contribution of surface tension to the chemical potential $\mu'$ must be considered.  The corresponding term $\sim \theta^2 \sigma w^2$ should be added to the right hand side of Eq$.$ (\ref{eq:mu'}).  This produces a smaller thickness $d$ at all humidities, and allows for finite $d$ at $100\%$ humidity \cite{Derjaguin}.

\section{Charge unbinding and conductivity}

We suppose that the surface contains some two-dimensional concentration $N$ of fixed charges, which for the sake of discussion we take to be negative (these may be, for example, the phosphate groups of DNA).  When the surface is dry, these negative charges tend to be neutralized by positive ions that are tightly-bound to the negatives by electrostatic attraction.  Depending on the method of surface preparation, the positive charges may be either protons H$^+$ or some metal ion like Na$^+$.  If the surface was obtained by drying from distilled water, it will most likely be covered by H$^+$ ions.  If the surface was exposed to salty water, then it will attract positive salt ions in order to maintain neutrality.  We assume that any excess salt has been rinsed out.

In order for electrical conduction to occur, some of these positive-negative pairs must unbind.  In the absence of adsorbed water, this unbinding process requires a huge activation energy $e^2/16 \pi \epsilon_0 a \approx 3.5$ eV.  Here, $a \approx 1$ \r{A} is the ion radius, $\epsilon_0$ is the vacuum permittivity, and the two charge species are taken to have charge $\pm e$.

The adsorption of a water layer surrounding a bound ion pair, however, lowers the activation energy by providing an atmosphere of high dielectric constant.  Electric field lines from a single unpaired ion remain preferrentially within the water layer, and as a result the electrostatic self-energy decreases.  In this section we estimate the activation energy associated with the unbinding of an ion pair immersed in a water layer of thickness $d$, and from there we calculate the density of free charges and the conductivity.

\begin{figure}[htb]
\centering
\includegraphics[width=0.45 \textwidth]{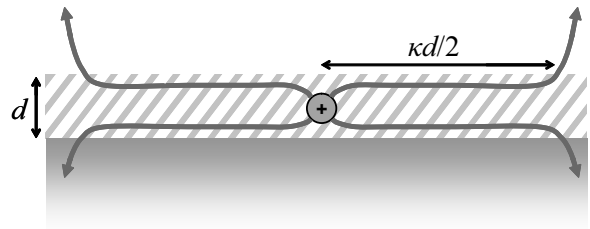}
\caption{Schematic depiction of electric field lines in the vicinity of an unpaired charge within a water layer (hatched area) that has condensed on top of a substrate (solid area).  Field lines remain preferrentially within the water layer, exiting only at a distance $L \simeq \kappa d/2$.} \label{fig:flat_layer}
\end{figure}

The Coulomb self-energy of a charge confined within a ``slab" geometry has been solved exactly in Refs$.$ \cite{Keldysh, Netz}.  Here we present a heuristic derivation to elucidate the structure of the electric field.  We imagine that a single unpaired charge produces electric field lines that remain inside the water layer, spreading out radially in two dimensions, until some distance $L$ when they exit the layer and behave three-dimensionally.  This is shown schematically in Fig$.$ \ref{fig:flat_layer}.  Under these assumptions it is easy to assemble a piecewise description of the electric field using Gauss's Law:
\be 
E(r) = \left\{
\begin{array}{c l}
  e/2 \pi \epsilon_0 \kappa r d, & d/2 < r < L\\
  e/4 \pi \epsilon_0 r^2, & r > L
\end{array}
\right. .
\ee
Here, $\kappa \approx 80$ is the dielectric constant of water, and the charge is taken to reside in the center of the water layer ($r = 0$).  It is straightforward to calculate the Coulomb self-energy $U$ of the charge:
\begin{eqnarray} 
U & = & \frac{\epsilon_0 \kappa}{2} \int E(r)^2 dV \nonumber \\
& = & \frac{e^2}{4 \pi \epsilon_0 \kappa d} \ln \left(2 L/d \right) + \frac{e^2}{8 \pi \epsilon_0 L} . \label{eq:U}
\end{eqnarray}

The distance $L$ at which electric field lines leave the film is that which minimizes the self-energy:
\be 
\frac{ \partial U}{\partial L} = 0.
\ee
Applying this condition to Eq$.$ (\ref{eq:U}) leads to the conclusion $L = \kappa d / 2$.  This length plays the role of the screening radius, truncating the two-dimensional logarithmic potential at the distance $r = \kappa d/2$.  As a result, the self-energy of a free charge in the water film can be estimated as
\be 
U \simeq \frac{e^2}{4 \pi \epsilon_0 \kappa d} (1 + \ln \kappa).
\ee

A more accurate estimate for $U$ can be obtained from the asymptotic expression for the interaction energy $V(r)$ of two charges $\pm e$ within the water film and separated a distance $r$ with $d \ll r \ll \kappa d$ \cite{Keldysh}:
\be 
V(r) \simeq -\frac{e^2}{2 \pi \epsilon_0 \kappa d} \left[ -\gamma + \ln \left( \kappa d/r \right) \right].
\ee
Here, $\gamma \approx 0.577$ is the Euler constant.  The self-energy of each unpaired ion can be estimated as $U = (V(\infty) - V(d))/2$, or
\be
U = \frac{e^2}{4 \pi \epsilon_0 \kappa d} (-\gamma + \ln \kappa).
\ee
Further calculations will use this expression for $U$.

In order to calculate the density of free charges at a given layer thickness $d$, we define the free energy of the solution when some number $n$ of the possible $N$ ion pairs in a unit area are unbound.  We use as the zero of free energy the case where none of the ion pairs on the surface, of total area $A$, are unbound.  Then, if each unbound positive ion occupies an area $\sim w^2$, the free energy $F$ per unit area can be estimated as 
\be 
F = 2 n U - n k_B T \ln(1/n w^2) - \frac{k_B T}{A} \ln \left[ NA \choose nA \right] \label{eq:F}.
\ee
The first term of Eq$.$ (\ref{eq:F}) refers to the electrostatic self-energy of all unpaired charges (and assumes that $n \ll \epsilon_0 \kappa k_B T / e^2 d$, so that there is no screening of the self-energy $U$ \cite{Minnhagen}).  The second term corresponds to the positional entropy of the free positive charges, and the third term represents the ``hole entropy" of the unpaired, fixed negative charges.  Here $ X \choose Y $ is used to represent the binomial coefficient ``$X$ choose $Y$".

We can use Stirling's approximation to simplify Eq$.$ (\ref{eq:F}) and then impose the equilibrium condition $\partial F / \partial n = 0$ in order to derive a relation between $n$ and $d$.  We arrive at
\be 
\frac{n^2 w^2}{N - n} = e^{-\frac{2 l_B}{d} (-\gamma + \ln \kappa)} , \label{eq:nsquared}
\ee
where $l_B = e^2 / 4 \pi \epsilon_0 \kappa k_B T$ is the Bjerrum length; $l_B  \approx 7.0$ \r{A} at room temperature in bulk water.  When $n \ll N$, Eq$.$ (\ref{eq:nsquared}) reduces to
\be 
n = \sqrt{N/w^2} e^{-\frac{l_B}{d} (-\gamma + \ln \kappa)} . \label{eq:n}
\ee

The two-dimensional conductivity $\sigma = j/E$, where $E$ is the strength of the applied electric field and $j$ is the current per unit width in the direction of the applied field, can be determined from the density $n$ of available charge carriers as
\be 
\sigma = n e \mu \label{eq:simplesigma}.
\ee
Here, $\mu$ is the electrical mobility of the charge carriers.  Substituting Eq$.$ (\ref{eq:n}) into Eq$.$ (\ref{eq:simplesigma}) gives 
\be 
\sigma = e \mu \sqrt{N/w^2} e^{-\frac{l_B}{d} (-\gamma + \ln \kappa)} . \label{eq:sigmad}
\ee

The dependence of the water layer thickness $d$ on humidity may be taken as an empirical relation for a given surface, or it may be assumed to result from Van der Waals forces as in Eq$.$ (\ref{eq:d-RH}).  In the latter case,
\begin{eqnarray}
\sigma & = & e \mu \sqrt{N/w^2} \times \\ \nonumber
& & \exp \left(-\frac{l_B}{w} \left(6 \pi k_B T \ln(p/p_0) / H \right)^{1/3} (-\gamma + \ln \kappa) \right) . \label{eq:final}
\end{eqnarray}
If $\kappa$ is considered to be a constant, then the dependence of conductivity on humidity implied in Eq$.$ (\ref{eq:final}) can be expressed in simplified form as
\be 
\log \sigma = \alpha - \beta \left(\log(p_0/p)\right)^{1/3}, \label{eq:simplified}
\ee
where $\alpha$ and $\beta$ are positive constants.

In general, however, the effective dielectric constant of the water layer depends on its thickness.  The value of $\kappa$ for a water film is known to decay exponentially with film thickness to its bulk value \cite{Stobbe, Pashley}:
\be 
\kappa \simeq 80 (1 - e^{-d/\lambda}) \label{eq:kappa},
\ee
where $\lambda$ is some decay length.  This has the effect of increasing the Bjerrum length $l_B \propto 1/\kappa$ at low humidities relative to its bulk value.  The behavior of the dielectric constant in the immediate vicinity of a DNA molecule is not well-known, so here we use the estimate $\lambda = 3$ \r{A}.

The mobility $\mu$ in general depends on the nature of the charge carriers in the system.  As an example, we can make an estimate for proton hopping between water molecules (the ``Grotthus mechanism").  Experimental data suggests a hopping time of about $1.5$ ps in bulk water \cite{Grotthus}.  Using a hopping length of $2.5$ \r{A} --- the hydrogen bond length between water and H$_3$O$^+$ --- gives a 2D diffusion constant $1 \times 10^{-4}$ cm$^2$/s and a corresponding mobility $4 \times 10^{-3}$ cm$^2$/Vs.  If the total density of fixed charges is $N = 10^{-2}$ \r{A}$^{-2}$, then we can estimate a maximum conductivity of about $10^{-7} \Omega^{-1}$.

\begin{figure}[htb]
\centering
\includegraphics[width=0.45 \textwidth]{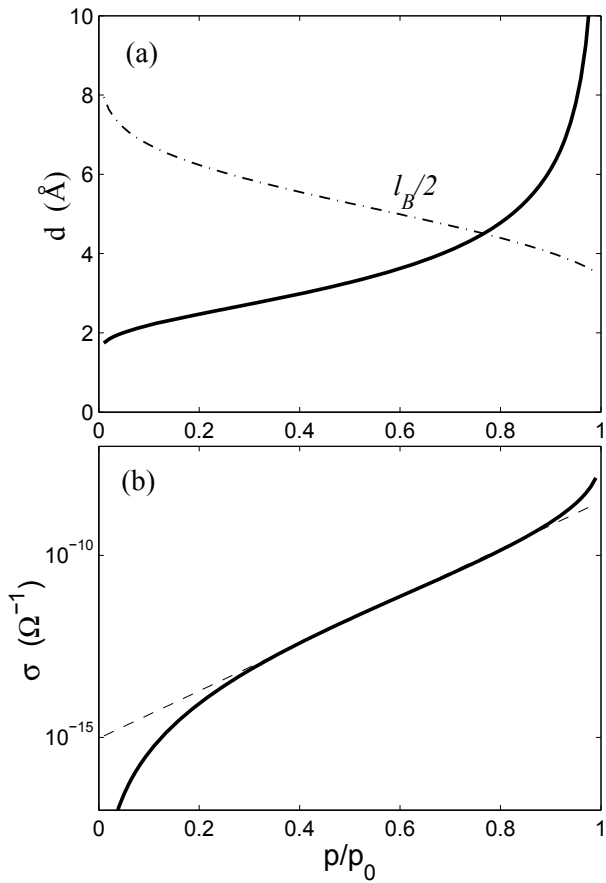}
\caption{(a) The water layer thickness as a function of humidity, from Eq$.$ (\ref{eq:d-RH}).  The dash-dotted line shows $l_B/2$, which decreases as the water layer becomes thicker and $\kappa$ approaches its bulk value.
\newline (b) Surface conductivity as a function of humidity (solid line), from Eq$.$ (\ref{eq:final}).  The maximum conductivity is $10^{-7} \Omega^{-1}$, as estimated in section III. The dotted line is a linear expansion around 50\% humidity.} \label{fig:thickness-conductivity}
\end{figure}

Fig$.$ \ref{fig:thickness-conductivity} shows a typical case for the thickness $d$ of the water layer, as given by Eq$.$ (\ref{eq:d-RH}),  and the conductivity as functions of humidity.  We have used changing a dielectric constant $\kappa$ as in Eq$.$ (\ref{eq:kappa}) and the estimated Hamaker constant $H = -0.4$ eV.

\section{Formation of a water cap on a free ion}

\begin{figure}[htb]
\centering
\includegraphics[width=0.45 \textwidth]{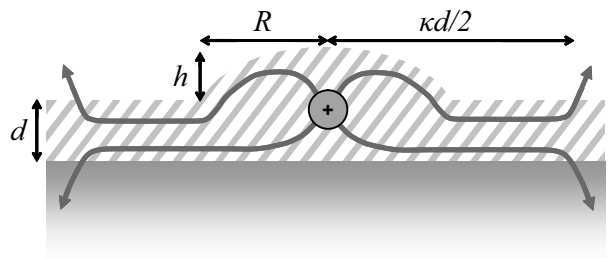}
\caption{Schematic illustration of a water cap around a free ion, formed by the adsorption of additional water molecules to the region of intense electric field surrounding the ion.  The radius $R$ and height $h$ of the cap can be optimized to determine the effect of the cap formation on the conductivity.} \label{fig:swollen_layer}
\end{figure}

To this point we have assumed that the water layer is flat, with a constant thickness $d$ determined solely by the competition between the free energy of condensation and the attraction of water molecules to the surface (Eqs$.$ (\ref{eq:Fads}) and (\ref{eq:mu'})).  In other words, we have treated the Coulomb energy as a small perturbation.  This is a good approximation for relatively thick films, where the Coulomb energy is small.  At humidities lower than about $40\%$, however, we are aparently dealing with water layers of single-molecule thickness $d \simeq w$.  Here the large Coulomb energy could force the water film to swell above a free ion, forming an additional cap of water as depicted schematically in Fig$.$ \ref{fig:swollen_layer}.

In this way the Coulomb field of free charges contributes to the disjoining pressure on the water film.  The creation of a water cap is resisted by the combination of free energies in Eqs$.$ (\ref{eq:Fads}) and (\ref{eq:mu'}), integrated over the number of water molecules that comprise the cap.  The surface tension energy associated with the ``bulge" in the water surface must also be taken into account.  We assume that the cap has some characteristic radius $R$ and height $h$ above the flat layer.  The free energy $\Delta F_{cap}$ associated with the formation of the cap depends on $R$ and $h$, and can be written most generally as
\be 
\Delta F_{cap} = \Delta U + \Delta F_{VdW} + \Delta F_{ads} + \Delta F_{surf} ,
\ee 
where $\Delta U$ is the (negative) change in electrostatic energy associated with the formation of the cap, $\Delta F_{VdW}$ is the (negative) Van der Waals energy resulting from adsorption of the water molecules comprising the cap, $\Delta F_{ads}$ is the (positive) condensation free energy associated with their adsorption, and $\Delta F_{surf}$ is the (positive) additional surface energy.  Optimizing $\Delta F_{cap}$ with respect to $R$ and $h$ determines the equilbrium cap size and its contribution to the free energy of an unbound charge.  For the example shown in Fig$.$ \ref{fig:thickness-conductivity} at 20\% humidity, we find that the equilibrium $\Delta F_{cap} \approx -0.4 k_BT$.  As a consequence, the conductivity is increased by only about 50\%, which in logarithmic scale would be a barely-noticeable correction to Fig$.$ \ref{fig:thickness-conductivity}.  

\section{Discussion}

The prediction for conductivity of Fig$.$ (\ref{fig:thickness-conductivity}) is in reasonable agreement with experiments \cite{Awakuni, Anderson, Sakai, Heim, Sadaoka, Seiyama, HanHa, Yamahata, Kleine-Ostmann, Tuukkanen, Otsuka, Matsuo, Armitage}: it reproduces the exponential dependence of conductivity on humidity, particularly in the well-studied range of humidities $p/p_0 > 0.3$, and suggests that the humidity can span six orders of magnitude for a realistic value of the Hamaker constant $H$.  However, for DNA our estimates yield a conductivity that is smaller than observed values by one or two orders of magnitude.  This may be the result of an unrealistically low estimate of the mobility based on data for bulk water.  Interactions between conducting ions and the negative surface charges may improve the mobility.

It should be emphasized that Eqs$.$ (\ref{eq:n}) and (\ref{eq:sigmad}) are only valid when free ions are exponentially rare, i.e. $U \gg k_BT$.  When the water layer becomes sufficiently thick, the self-energy of a free charge becomes comparable to $k_BT$ and the system undergoes a Kosterlitz-Thouless (KT) unbinding transition at $d = l_B/2$ \cite{Minnhagen}.  Strictly speaking, the formulas derived in this work are only valid below the transition, $d < l_B/2$, and immediately after it at $2d/l_B - 1 < (\ln \kappa)^{-2}$ (see Fig$.$ 2 in Ref$.$ \cite{Minnhagen}).  For larger $d$, we expect a rapid acceleration in the density of free charges $n$.  For $d > l_B$, $n$ becomes saturated, so that the conductivity no longer depends strongly on humidity.  In the example of Fig$.$ \ref{fig:thickness-conductivity}, we expect the KT transition to occur at about 80\% humidity, and the corresponding steep rise of $n$ should become apparent at about 85\%.  We are not aware of any experimental evidence which demonstrates this behavior.

For conduction through an assembly of DNA molecules, we expect that the conductivity should depend in general on the geometry of the assembly.  In the case of a single double-helical DNA molecule with an adsorbed water layer, the electric field from an unpaired charge behaves one-dimensionally at distances larger than half the circumference of the double-helix \cite{JZhang, Burin}. As a result the electrostatic self-energy is larger than our prediction and the conductivity should be smaller.  For a bundle of parallel double-helices, if water is absorbed within the bundle then the electric field of a free charge exhibits three-dimensional behavior at distances larger than the interhelical spacing. The result is a smaller self-energy and a larger resulting conductivity.  Surprisingly, experiments on individual DNA molecules \cite{Tuukkanen}, DNA films \cite{Kleine-Ostmann, Otsuka, Matsuo, Armitage}, and DNA bundles \cite{HanHa, Yamahata} all show a similar exponential dependence of conductivity on humidity.  The lack of a geometry-dependent difference remains a puzzle.

The temperature dependence of DNA conductivity was studied only by the authors of Ref$.$ \cite{HanHa} at about 50\% humidity.  They obtained an activation-like dependence $\sigma \propto e^{-E_a/k_BT}$ with $E_a \approx 0.5$ eV.  We cannot account for this behavior because of the temperature dependence of the dielectric constant, which in bulk water has the dependency $\kappa \propto T^{-1.44}$ \cite{CRC}.  Further experimental work examining the temperature dependence of DNA conductivity could help improve our understanding.

Finally, this paper has considered only the case where fixed charges on the conducting surface are of a single sign, \textit{e.g.} fixed negative charges with positive counterions.  One may well ask how much conductivity should be expected in the case of a charge-neutral surface, where both positive and negative fixed charges are present on the surface with their respective counterions.  If the density of fixed charges $N$ on the surface is such that fixed charges are very sparse, then there is no interaction between fixed charges and we expect that the predictions of this paper should remain valid.  Each fixed charge may lose its counterion via the unbinding process described in this paper, and free ions contribute equally to the conductivity regardless of sign.  If $N$ is large, however, then fixed charges of opposite sign may neutralize each other and allow their respective counterions to unbind with very little energy cost.  If this is the case, then counterions may be rinsed out during the surface preparation, resulting in a depletion of the conductivity.  A small amount of conduction may still occur even in the absence of counterions through the ionization of water molecules, but the characterization of this mechanism is beyond the scope of the present work.

\begin{small}
\vspace*{2ex} \par \noindent
{\em Acknowledgements.}

We are grateful to A. Burin, E. Brodskaya, A. Shchekin, M. Fogler, and A. Ankudinov for helpful discussions.  Brian Skinner acknowledges the support of the NSF Graduate Research Fellowship.  M.S. Loth thanks the Fine Theoretical Physics Institute for financial support.
\end{small}



\begin{thebibliography}{100}
\bibitem{Hfootnote} The Hamaker constant for the water-adsorbent interaction is defined as $H = H_{w,w} - H_{s,w}$, where $H_{w,w}$ is the Hamaker constant of two masses of water interacting through vacuum and $H_{s,w}$ corresponds to a mass of water interacting through vacuum with a mass of the adsorbent material.
\end{thebibliography}
\end{document}